\PassOptionsToPackage{unicode}{hyperref}
\PassOptionsToPackage{hyphens}{url}
\documentclass[
  10pt,
]{article}
\usepackage{amsmath,amssymb}
\usepackage{iftex}
\ifPDFTeX
  \usepackage[T1]{fontenc}
  \usepackage[utf8]{inputenc}
  \usepackage{textcomp} 
\else 
  \usepackage{unicode-math} 
  \defaultfontfeatures{Scale=MatchLowercase}
  \defaultfontfeatures[\rmfamily]{Ligatures=TeX,Scale=1}
\fi
\usepackage{lmodern}
\ifPDFTeX\else
\fi
\IfFileExists{upquote.sty}{\usepackage{upquote}}{}
\IfFileExists{microtype.sty}{
  \usepackage[]{microtype}
  \UseMicrotypeSet[protrusion]{basicmath} 
}{}
\makeatletter
\@ifundefined{KOMAClassName}{
  \IfFileExists{parskip.sty}{%
    \usepackage{parskip}
  }{
    \setlength{\parindent}{0pt}
    \setlength{\parskip}{6pt plus 2pt minus 1pt}}
}{
  \KOMAoptions{parskip=half}}
\makeatother
\usepackage{xcolor}
\usepackage{color}
\usepackage{fancyvrb}

\DefineVerbatimEnvironment{Highlighting}{Verbatim}{commandchars=\\\{\}}
\newenvironment{Shaded}{}{}

\newcommand{\AttributeTok}[1]{\textcolor[rgb]{0.49,0.56,0.16}{#1}}

\newcommand{\CommentTok}[1]{\textcolor[rgb]{0.38,0.63,0.69}{\textit{#1}}}

\newcommand{\DataTypeTok}[1]{\textcolor[rgb]{0.56,0.13,0.00}{#1}}

\newcommand{\ExtensionTok}[1]{#1}

\newcommand{\ImportTok}[1]{\textcolor[rgb]{0.00,0.50,0.00}{\textbf{#1}}}

\newcommand{\NormalTok}[1]{#1}
\newcommand{\OperatorTok}[1]{\textcolor[rgb]{0.40,0.40,0.40}{#1}}

\newcommand{\StringTok}[1]{\textcolor[rgb]{0.25,0.44,0.63}{#1}}

\usepackage{longtable,booktabs,array}
\usepackage{calc} 
\usepackage{etoolbox}
\makeatletter
\patchcmd\longtable{\par}{\if@noskipsec\mbox{}\fi\par}{}{}
\makeatother
\IfFileExists{footnotehyper.sty}{\usepackage{footnotehyper}}{\usepackage{footnote}}
\makesavenoteenv{longtable}
\providecommand{\tightlist}{%
  \setlength{\itemsep}{0pt}\setlength{\parskip}{0pt}}
\usepackage[margin=1in]{geometry}
\usepackage{mathptmx}
\ifLuaTeX
  \usepackage{selnolig}  
\fi
\IfFileExists{bookmark.sty}{\usepackage{bookmark}}{\usepackage{hyperref}}
\IfFileExists{xurl.sty}{\usepackage{xurl}}{} 
\hypersetup{
  pdftitle={AIPC: Agent-Based Automation for AI Model Deployment with Qualcomm AI Runtime},
  hidelinks,
  pdfcreator={LaTeX via pandoc}}

\title{AIPC: Agent-Based Automation for AI Model Deployment with
Qualcomm AI Runtime}
\author{
Jianhao Su, Zhanwei Wu, ShengTing Huang, Weidong Feng\\
Qualcomm Technologies, Inc.
}
\date{April 2026}

\begin{document}
\maketitle
\begin{abstract}
Edge AI model deployment is a multi-stage engineering process involving
model conversion, operator compatibility handling, quantization
calibration, runtime integration, and accuracy validation. In practice,
this workflow is long, failure-prone, and heavily dependent on
deployment expertise, particularly when targeting hardware-specific
inference runtimes. This technical report presents AIPC (AI Porting
Conversion), an AI agent-driven approach for constrained automation of
AI model deployment. AIPC decomposes deployment into standardized,
verifiable stages and injects deployment-domain knowledge into agent
execution through Agent Skills, helper scripts, and a stage-wise
validation loop. This design reduces both the expertise barrier and the
engineering time required for hardware deployment.

Using Qualcomm AI Runtime (QAIRT) as the primary scenario, this report
examines automated deployment across representative vision, multimodal,
and speech models. In the cases covered here, AIPC can complete
deployment from PyTorch to runnable QNN/SNPE inference within
\textbf{7-20 minutes} for structurally regular vision models, with
indicative API costs roughly in the range of USD \textbf{0.7-10}. For
more complex models involving less-supported operators, dynamic shapes,
or autoregressive decoding structures, fully automated deployment may
still require further advances, but AIPC already provides practical
support for execution, failure localization, and bounded repair.
\end{abstract}

\textbf{Keywords}: AI model deployment; edge AI; QAIRT; large language
model agent; automated software engineering; model conversion

\textbf{Open-source repository}:
https://github.com/quic/ai-engine-direct-helper/tree/main/tools/skills/qai-runner-skill

\hypertarget{introduction}{%
\section{Introduction}\label{introduction}}

\hypertarget{background}{%
\subsection{Background}\label{background}}

As edge AI applications expand from image classification to object
detection, speech recognition, multimodal understanding, and generative
tasks, model deployment is no longer a secondary step after training. It
has become a key engineering stage that often determines whether a
system can be delivered in practice, especially in scenarios requiring
low latency, privacy preservation, and offline availability.

However, a substantial gap remains between research-oriented model code
and reliable execution on edge runtimes. In practice, deployment
involves far more than model export: it often requires ONNX conversion,
operator replacement, dynamic-shape resolution, graph repair, runtime
integration, preprocessing and postprocessing adaptation, accuracy
validation, and hardware-specific compilation. For a hardware-optimized
runtime such as Qualcomm AI Runtime (QAIRT), developers must understand
not only the model itself, but also toolchain behavior, operator-support
boundaries, layout constraints, and platform-specific differences.
Consequently, deployment remains a knowledge-intensive and
experience-dependent task.

\hypertarget{research-question}{%
\subsection{Research Question}\label{research-question}}

Large language models (LLMs) have already shown strong capabilities in
code generation, automated repair, task orchestration, and tool usage.
Based on this trend, this paper studies the following question:

\begin{quote}
Given an existing deployment toolchain and validation workflow, can an
LLM agent effectively take over repetitive engineering work in AI model
deployment, and perform bounded automatic repair when conversion fails
or runtime exceptions occur?
\end{quote}

The key issue is not whether ``an LLM knows every hardware detail,'' but
rather whether, when domain knowledge is explicitly provided through
Agent Skills, helper functions, and validation rules, the LLM can act as
a task orchestrator and executor that turns a complex deployment process
into a repeatable, verifiable, and continuously improvable automated
workflow.

\hypertarget{contributions}{%
\subsection{Contributions}\label{contributions}}

The main contributions of this paper are summarized as follows:

\begin{enumerate}
\def\labelenumi{\arabic{enumi}.}
\tightlist
\item
  \textbf{Agentizing the AI deployment workflow}: We reorganize existing
  deployment workflows into a form executable by AI agents. Through
  Agent Skills, a validation loop, and failure-recovery mechanisms, a
  deployment pipeline that traditionally depends on human experience is
  transformed into a repeatable and verifiable automated workflow, with
  complete end-to-end QAIRT-based cases presented in this paper.
\item
  \textbf{Introducing the ``Skills + validation loop'' design pattern}:
  Toolchain knowledge is encapsulated through Agent Skills and combined
  with golden outputs, interface checks, and consistency comparisons to
  constrain agent behavior within a verifiable space.
\item
  \textbf{Providing a repair-level view of deployment-oriented model
  surgery}: We organize deployment repair actions across the PyTorch
  source level, ONNX graph level, and runtime interface level, enabling
  compatibility fixes to be incorporated into an automated workflow.
\item
  \textbf{Conducting representative multi-model, multi-agent case
  analysis}: Across models of varying structural complexity and several
  mainstream AI agents, we summarize success boundaries for automation,
  manual intervention patterns, and differences in agent behavior.
\item
  \textbf{Deriving engineering practice lessons}: We extract a set of
  design principles suitable for edge AI deployment automation and
  provide guidance for future extension to other inference frameworks.
\end{enumerate}

\hypertarget{positioning-as-a-technical-report}{%
\subsection{Positioning as a Technical
Report}\label{positioning-as-a-technical-report}}

This paper is written as a technical report. It focuses on system
design, engineering workflow, case observations, and methodological
lessons, rather than emphasizing large-scale statistically significant
experiments under strictly controlled conditions. The goal is to report
engineering evidence and reusable workflow design, rather than to claim
benchmark superiority under universally controlled conditions. The
experimental results are primarily used to support the following
claims:\\
(1) AIPC is engineering-feasible as a workflow framework;\\
(2) different model categories exhibit structurally different
difficulties in automated deployment;\\
(3) the value of LLM agents in deployment tasks is better understood as
that of a ``constrained automation executor'' rather than a fully
autonomous end-to-end solver.

\hypertarget{paper-organization}{%
\subsection{Paper Organization}\label{paper-organization}}

The rest of the paper is organized as follows. Section 2 reviews related
work. Section 3 introduces the QAIRT background and deployment
challenges. Section 4 presents the AIPC method design. Section 5
describes the evaluation targets and procedure. Section 6 reports case
results and observations. Section 7 summarizes engineering lessons and
limitations. Section 8 discusses future directions. Section 9 concludes
the paper.

\hypertarget{open-source-release}{%
\subsection{Open-Source Release}\label{open-source-release}}

The workflow and supporting materials described in this paper have been
released in open-source form as part of the QAI AppBuilder toolchain
since \textbf{March 10, 2026}. This release is intended to provide users
with a practical starting point for constrained, verifiable AI
deployment automation under QAIRT-centered workflows, including Agent
Skills, helper scripts, and related references. The goal of the release
is not to present a fully autonomous universal deployment system, but to
make the AIPC method reusable, inspectable, and continuously improvable
through open engineering practice.

\hypertarget{related-work}{%
\section{Related Work}\label{related-work}}

\hypertarget{ai-model-deployment-and-inference-frameworks}{%
\subsection{AI Model Deployment and Inference
Frameworks}\label{ai-model-deployment-and-inference-frameworks}}

Current mainstream AI deployment frameworks can be roughly divided into
several categories.

\textbf{Cross-platform inference frameworks}: LiteRT (formerly
TensorFlow Lite), ONNX Runtime, and MNN emphasize unified model
representation and multi-platform adaptability. These frameworks offer
good portability for general inference, but deep optimization for
specific hardware accelerators often still requires extra engineering
effort, and they typically do not provide automated orchestration for
the deployment workflow itself.

\textbf{Hardware-specific inference frameworks}: NVIDIA TensorRT
{[}1{]}, Intel OpenVINO {[}2{]}, Qualcomm QAIRT {[}8{]}, and Rockchip
RKNPU {[}3{]} usually provide higher performance and lower latency on
specific hardware, but also introduce stricter operator-support
boundaries, data-layout constraints, and toolchain dependencies. Their
deployment workflows depend heavily on engineers' understanding of
platform details and generally lack systematic automation support.

\textbf{Compiler-driven inference frameworks and
intermediate-representation infrastructure}: Apache TVM {[}22{]} is an
end-to-end deep learning compiler framework for multiple hardware
backends. Through a unified computational graph representation and
auto-tuning mechanisms (AutoTVM/Ansor), it compiles models into
executable code optimized for target hardware and offers strong
cross-platform generalization. MLIR (Multi-Level Intermediate
Representation) {[}23{]}, proposed by the LLVM community, is a general
compiler infrastructure that provides a unified framework for IR
transformations at different levels, and it has been widely used to
build ML compiler backends such as TensorFlow XLA and IREE. The
advantage of such frameworks is that they turn hardware adaptation into
the composition and optimization of compiler passes, giving them strong
extensibility. However, their deployment workflows still require
engineers to understand compiler abstraction layers and target hardware
characteristics in depth, and tuning for specific hardware often needs
either extensive manual configuration or costly automated search.

Overall, existing systems focus more on ``how to convert models better''
than on ``how to systematize deployment itself into an automated
workflow executable by agents.''

\hypertarget{llm-assisted-software-engineering-and-agentic-execution}{%
\subsection{LLM-Assisted Software Engineering and Agentic
Execution}\label{llm-assisted-software-engineering-and-agentic-execution}}

In recent years, tools such as GitHub Copilot {[}5{]}, Claude Code {[}4,
6{]}, Cursor {[}7{]}, OpenCode {[}18{]}, Cline {[}19{]}, and Codex CLI
{[}20{]} have promoted agentic use of LLMs in software engineering.
Through tool interfaces for file operations, command execution, and
context awareness, these systems extend LLMs from pure text generators
into ``software agents'' capable of executing tasks.

However, progress on general programming tasks does not imply direct
transferability to AI deployment. Deployment tasks have the following
characteristics:

\begin{itemize}
\tightlist
\item
  They depend on extensive \textbf{domain-specific toolchain knowledge}.
\item
  Correctness must be determined by \textbf{runtime result validation},
  not surface-level code appearance.
\item
  Failures often arise from interactions among \textbf{environment,
  framework, operator, and platform constraints}.
\item
  Many repairs require repeated localization across \textbf{PyTorch
  {[}10{]}, ONNX {[}9{]}, and runtime interfaces}.
\end{itemize}

Therefore, if domain knowledge is not injected into the agent context in
a structured form, and one relies only on the generic prior knowledge of
an LLM, common failures include hallucinated APIs, incorrect choice of
repair layer, and incorrect environment diagnosis.

\hypertarget{how-this-work-differs}{%
\subsection{How This Work Differs}\label{how-this-work-differs}}

Compared with existing deployment tools or LLM coding assistants, AIPC
is distinguished not by a new low-level compiler, but by an
\textbf{upper-layer workflow design for deployment automation}. Its main
differences are:

\begin{itemize}
\tightlist
\item
  \textbf{Explicit knowledge injection}: Platform knowledge is turned
  into reusable execution templates through Agent Skills and reference
  documents.
\item
  \textbf{Stage-wise verification}: Each stage is tied to observable
  inputs and outputs whenever possible, preventing the agent from
  drifting in unverifiable states.
\item
  \textbf{Failure recoverability}: Model surgery, retries, and human
  correction are integrated into the workflow, rather than treating
  failure as the end of the process.
\item
  \textbf{Engineering reusability}: The method emphasizes cross-model
  reuse, error accumulation, and potential for extension to other
  deployment tasks.
\end{itemize}

\hypertarget{background-common-challenges-in-hardware-specific-inference-deployment---qairt-as-an-example}{%
\section{Background: Common Challenges in Hardware-Specific Inference
Deployment - QAIRT as an
Example}\label{background-common-challenges-in-hardware-specific-inference-deployment---qairt-as-an-example}}

\hypertarget{overview-of-the-qairt-deployment-pipeline}{%
\subsection{Overview of the QAIRT Deployment
Pipeline}\label{overview-of-the-qairt-deployment-pipeline}}

Qualcomm AI Runtime (QAIRT) is the official AI inference SDK for
Qualcomm hardware platforms, covering the full range of inference
scenarios from lightweight vision models to generative large models. The
main forms involved in this paper include:

\begin{itemize}
\tightlist
\item
  \textbf{SNPE (Snapdragon Neural Processing Engine)}: A higher-level
  runtime framework that uses the DLC (Deep Learning Container) format.
  It provides good cross-backend execution capability and is suitable
  for rapid integration and prototyping.
\item
  \textbf{QNN (Qualcomm Neural Network)}: A framework closer to the
  hardware execution layer. It typically achieves high-performance
  deployment by generating platform-specific libraries and Context
  Binaries, and is the preferred path for production.
\item
  \textbf{Genie (Generative AI Inference Extension)}: An inference
  extension optimized for generative AI workloads. It provides efficient
  inference support on Qualcomm NPUs for large language models,
  multimodal models, and related tasks.
\end{itemize}

\textbf{QAI AppBuilder} provides a higher-level interface on top of
QAIRT, allowing developers to integrate model inference quickly in
Python and C++, and also offering OpenAI API-compatible interfaces for
large-model applications.

In the typical scenario considered in this paper, the deployment
pipeline can be summarized as:

\textbf{PyTorch -\textgreater{} ONNX -\textgreater{} QAIRT (SNPE or QNN)
-\textgreater{} Context Binary -\textgreater{} Runtime Inference}

This pipeline involves not only model format conversion, but also
interface alignment, runtime adaptation, and accuracy checking.

\hypertarget{typical-deployment-steps}{%
\subsection{Typical Deployment Steps}\label{typical-deployment-steps}}

\begin{itemize}
\tightlist
\item
  \textbf{Exporting PyTorch to ONNX}: Use methods such as
  \texttt{torch.onnx.export()} in PyTorch {[}10{]} to export a model to
  ONNX {[}9{]}. This stage must handle dynamic axes, input-output
  definitions, export versions, and differences in operator semantics.
\item
  \textbf{Converting ONNX to QAIRT}: Call \texttt{qnn-onnx-converter} or
  related tools to generate an intermediate representation consumable by
  QNN/SNPE. Common issues at this stage include unsupported operators,
  layout rewrites, shape-inference failures, and tool errors.
\item
  \textbf{Context compilation (AOT)}: Use Ahead-of-Time (AOT)
  compilation to generate a Context Binary for a specific target
  platform. Compared with runtime Just-in-Time (JIT) compilation, AOT
  significantly reduces model loading and initialization overhead and
  performs hardware-specific optimization in advance, improving
  edge-side execution efficiency.
\item
  \textbf{Quantization and calibration}: In performance-sensitive
  scenarios, further FP16, W8A16, or W8A8 quantization may be applied.
  This step depends heavily on representative calibration data and a
  stable preprocessing pipeline.
\item
  \textbf{Runtime integration and validation}: Load the model through
  the application-layer interface, run sample inputs, and compare
  outputs with the reference baseline to confirm that the converted
  result is usable.
\end{itemize}

\hypertarget{analysis-of-deployment-difficulties}{%
\subsection{Analysis of Deployment
Difficulties}\label{analysis-of-deployment-difficulties}}

\hypertarget{operator-support-boundaries}{%
\subsubsection{Operator support
boundaries}\label{operator-support-boundaries}}

QAIRT supports only a subset of ONNX operators. When unsupported
operators appear in a model, developers often need to perform equivalent
replacement or graph rewriting before deployment can proceed.

\hypertarget{data-layout-changes}{%
\subsubsection{Data layout changes}\label{data-layout-changes}}

Conversion tools may automatically change tensor layouts during
intermediate stages, causing original preprocessing and postprocessing
logic to fail.

\hypertarget{dynamic-shape-and-fixed-shape-conflicts}{%
\subsubsection{Dynamic-shape and fixed-shape
conflicts}\label{dynamic-shape-and-fixed-shape-conflicts}}

Many research models naturally include dynamic input lengths, dynamic
batch sizes, or dynamic sequence structures, while edge inference
frameworks generally prefer fixed-shape compilation paths.

\hypertarget{quantization-depends-on-data-and-workflow-stability}{%
\subsubsection{Quantization depends on data and workflow
stability}\label{quantization-depends-on-data-and-workflow-stability}}

Quantization is not a step that can be completed simply by issuing a
command. If calibration data deviates from the true input distribution,
or if preprocessing logic is inconsistent with the actual inference
pipeline, quantization results may suffer severe accuracy degradation.

\hypertarget{platform-and-environment-complexity}{%
\subsubsection{Platform and environment
complexity}\label{platform-and-environment-complexity}}

In environments involving Windows on ARM, x86 emulation, and multiple
coexisting toolchains, an agent may misjudge the architecture, invoke
the wrong library version, or misunderstand the context of a tool error.
These issues are not caused by the model itself, but by complex
interactions at the deployment-environment level.

\hypertarget{the-aipc-method}{%
\section{The AIPC Method}\label{the-aipc-method}}

\hypertarget{overall-design-philosophy}{%
\subsection{Overall Design Philosophy}\label{overall-design-philosophy}}

AIPC (AI Porting Conversion) is not a single script or point tool, but
an automation organization method for deployment tasks. Although its
methodology has a degree of generality, the system implementation, step
design, and experimental validation in this paper are all centered on
QAIRT. Its core ideas are:

\begin{itemize}
\tightlist
\item
  Break a complex deployment pipeline into several \textbf{verifiable
  stages}.
\item
  Solidify hardware deployment knowledge into \textbf{reusable Agent
  Skills and reference templates}.
\item
  Let AI agents perform code generation, command invocation, error
  localization, and local repair inside a \textbf{constrained workflow}.
\item
  Accumulate automation failures into reusable knowledge for future
  tasks.
\end{itemize}

In other words, the point of AIPC is not to ``let the agent improvise
freely,'' but to ``let the agent work inside an engineered space.''

\hypertarget{workflow-decomposition}{%
\subsection{Workflow Decomposition}\label{workflow-decomposition}}

AIPC divides the deployment process into six major steps, as shown in
Figure 1.

\begin{samepage}
\begin{Shaded}
\begin{Highlighting}[]
\NormalTok{[Figure 1: Conceptual Diagram of the AIPC Workflow]}

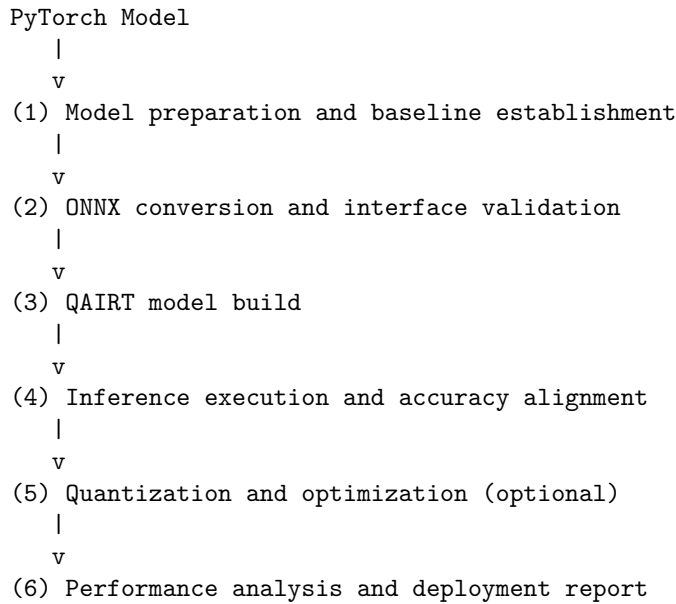
\NormalTok{PyTorch Model}
\NormalTok{   |}
\NormalTok{   v}
\NormalTok{(1) Model preparation and baseline establishment}
\NormalTok{   |}
\NormalTok{   v}
\NormalTok{(2) ONNX conversion and interface validation}
\NormalTok{   |}
\NormalTok{   v}
\NormalTok{(3) QAIRT model build}
\NormalTok{   |}
\NormalTok{   v}
\NormalTok{(4) Inference execution and accuracy alignment}
\NormalTok{   |}
\NormalTok{   v}
\NormalTok{(5) Quantization and optimization (optional)}
\NormalTok{   |}
\NormalTok{   v}
\NormalTok{(6) Performance analysis and deployment report}
\end{Highlighting}
\end{Shaded}
\end{samepage}

\hypertarget{step-1-model-preparation-and-baseline-establishment}{%
\subsection{Step 1: Model Preparation and Baseline
Establishment}\label{step-1-model-preparation-and-baseline-establishment}}

The goal of this step is to confirm that the original PyTorch model runs
correctly in the current environment, and to establish the ``golden
reference'' used for subsequent validation.

Concretely, the agent needs to load the model and required weights,
determine the input specification, data types, and preprocessing logic,
use sample inputs to generate reference outputs, and save key
intermediate results, input-output shapes, and baseline data.

The significance of this stage is that every later conversion, repair,
or quantization step must return to this baseline to determine whether
the result is still trustworthy.

\hypertarget{step-2-onnx-conversion-and-modular-inference-pipeline}{%
\subsection{Step 2: ONNX Conversion and Modular Inference
Pipeline}\label{step-2-onnx-conversion-and-modular-inference-pipeline}}

At this stage, the agent typically needs to automatically generate an
ONNX export script and set the correct inputs, outputs, opset, and
dynamic axes. It also needs to split the original inference flow into
three independent modules: \textbf{preprocessing}, \textbf{model
inference}, and \textbf{postprocessing}. In addition, it must write ONNX
inference test code and numerically compare ONNX outputs with the
PyTorch baseline.

The core value of this modularization is that it decouples preprocessing
and postprocessing from the model body, making problem localization in
later conversion stages more precise while also providing a standardized
interface for large-scale generation of quantization calibration data.

\hypertarget{model-surgery}{%
\subsubsection{Model Surgery}\label{model-surgery}}

AIPC treats model surgery as a core capability in deployment automation
rather than as an exceptional case. Model surgery refers to
deployment-oriented rewriting of model structure under the premise of
preserving functional equivalence or approximate equivalence. The
surgeries involved in this paper include:

\begin{itemize}
\tightlist
\item
  Converting dynamic shapes to fixed shapes.
\item
  Replacing unsupported operators with equivalent operator compositions.
\end{itemize}

Model surgery is triggered by specific failure events in the deployment
pipeline, corresponding to three intervention levels:

\begin{itemize}
\tightlist
\item
  \textbf{When ONNX export fails}: modify module implementations at the
  PyTorch source level to eliminate operator or shape incompatibilities
  during export.
\item
  \textbf{When QAIRT conversion fails}: patch nodes and attributes at
  the ONNX graph level so that the conversion tool can process the model
  correctly.
\item
  \textbf{When QAIRT runtime fails} (including Context Binary
  compilation failure or inference execution exceptions): apply further
  compatibility fixes based on the actual error, and, if necessary, fall
  back to an earlier stage for repair.
\end{itemize}

\hypertarget{step-3-qairt-model-build}{%
\subsection{Step 3: QAIRT Model Build}\label{step-3-qairt-model-build}}

After the ONNX model passes validation, AIPC further converts it into
the QAIRT target format. The following strategies are especially
important at this stage:

\hypertarget{interface-preservation}{%
\subsubsection{Interface Preservation}\label{interface-preservation}}

To reduce interface inconsistency caused by layout drift, AIPC prefers
conservative strategies such as \texttt{-\/-preserve\_io} to keep model
inputs and outputs aligned with the ONNX side. For example:

\begin{Shaded}
\begin{Highlighting}[]
\ExtensionTok{qnn{-}onnx{-}converter} \AttributeTok{{-}{-}input\_network}\NormalTok{ model.onnx }\DataTypeTok{\textbackslash{}}
                   \AttributeTok{{-}{-}output\_path}\NormalTok{ model.cpp }\DataTypeTok{\textbackslash{}}
                   \AttributeTok{{-}{-}preserve\_io}
\end{Highlighting}
\end{Shaded}

In addition, \texttt{-\/-preserve\_io} preserves the input and output
ordering of the ONNX model as-is, avoiding application-layer interface
mismatches caused by internal tensor reordering in the conversion tool.

Although this strategy may reduce performance in some cases,
prioritizing interface stability is usually more beneficial for
successfully completing the full automated pipeline.

However, \texttt{-\/-preserve\_io} can be problematic in some
quantization pipelines: on certain SoC configurations it may implicitly
preserve FP I/O or trigger FP16-related constraints, which can surface
as platform errors (e.g., SocModel FP16 not supported) during quantized
model conversion or subsequent compilation. In such cases, a practical
workaround is to avoid the QNN quantization path and use the SNPE DLC
quantization flow instead, or adjust conversion options to not preserve
FP I/O depending on the target/runtime requirements.

\hypertarget{precision-configuration-and-target-hardware-adaptation}{%
\subsubsection{Precision Configuration and Target Hardware
Adaptation}\label{precision-configuration-and-target-hardware-adaptation}}

For models suitable for NPU execution, AIPC tends to prioritize FP16
precision mode at this stage because it is easier to validate for
accuracy and does not require calibration data. Some hardware platforms
may not natively support FP16 execution and may instead rely mainly on
INT8 or other quantized paths. In such cases, the corresponding
deployment artifacts are better handled in the later quantization stage,
rather than introducing quantization-related uncertainty during early
validation.

\hypertarget{automatic-context-binary-generation}{%
\subsubsection{Automatic Context Binary
Generation}\label{automatic-context-binary-generation}}

For the target platform, the agent further triggers Context Binary
compilation to produce an execution artifact closer to the final
deployable form.

\hypertarget{failure-recovery}{%
\subsubsection{Failure Recovery}\label{failure-recovery}}

If unsupported operators, shape issues, or toolchain exceptions appear
during conversion, the agent selects repairs at either the PyTorch level
or the ONNX level according to Skill rules and reruns the conversion,
instead of directly terminating the process.

\hypertarget{step-4-inference-execution-and-accuracy-alignment}{%
\subsection{Step 4: Inference Execution and Accuracy
Alignment}\label{step-4-inference-execution-and-accuracy-alignment}}

To reduce the complexity of integrating QAIRT inference for the agent,
AIPC primarily uses an AIPC loader approach so that an existing ONNX
inference script can be executed with the QAIRT backend without
modifying the original ONNX-side program:

\begin{Shaded}
\begin{Highlighting}[]
\CommentTok{\# Launch the inference script through the AIPC hot{-}patch loader,}
\CommentTok{\# which redirects the ONNX Runtime import to onnxwrapper.}
\ExtensionTok{python}\NormalTok{ aipc inference.py}
\end{Highlighting}
\end{Shaded}

For users who prefer explicit control, the same backend redirection can
also be written directly in code through the wrapper interface:

\begin{Shaded}
\begin{Highlighting}[]
\CommentTok{\# ONNX inference}
\ImportTok{import}\NormalTok{ onnxruntime }\ImportTok{as}\NormalTok{ ort}
\NormalTok{session }\OperatorTok{=}\NormalTok{ ort.InferenceSession(}\StringTok{"model.onnx"}\NormalTok{)}

\CommentTok{\# Explicit wrapper{-}based redirection to the QAIRT backend}
\ImportTok{import}\NormalTok{ onnxwrapper }\ImportTok{as}\NormalTok{ ort}
\NormalTok{session }\OperatorTok{=}\NormalTok{ ort.InferenceSession(}\StringTok{"model.onnx"}\NormalTok{, backend}\OperatorTok{=}\StringTok{"QAIRT"}\NormalTok{)}
\end{Highlighting}
\end{Shaded}

This design offers three benefits. First, it \textbf{reduces API
hallucination}, because the agent does not need to learn an entirely
unfamiliar inference interface. Second, it \textbf{facilitates alignment
verification}, since the same script can run under different backends.
Third, it \textbf{limits the scope of changes}, keeping fault
localization focused on differences between the model and the runtime
rather than on application-layer rewrites.

At this stage, AIPC typically verifies: - Whether inference completes
successfully. - Whether output shapes and data types match expectations.
- Whether errors relative to the ONNX or PyTorch baseline remain within
an acceptable range. - Whether preprocessing and postprocessing still
work correctly.

\hypertarget{step-5-quantization-and-optimization}{%
\subsection{Step 5: Quantization and
Optimization}\label{step-5-quantization-and-optimization}}

Only after the previous stages are stable does AIPC enter the
quantization stage. Its automation logic mainly relies on the fact that
the preprocessing module has already been standardized, enabling the
agent to generate calibration samples in batches and drive the
quantization toolchain.

In current practice, quantization mainly targets W8A16 and W8A8/INT8
configurations.

The core value of AIPC in this stage is not merely issuing a
quantization command, but organizing ``calibration data generation
-\textgreater{} quantization -\textgreater{} result validation'' into a
repeatable process. In the future, it can be further integrated with
toolsets such as AIMET to extend toward pruning and structured
compression.

\hypertarget{step-6-performance-analysis}{%
\subsection{Step 6: Performance
Analysis}\label{step-6-performance-analysis}}

Once the model runs stably, AIPC can continue by invoking tools such as
\texttt{qnn-net-run} and \texttt{snpe-net-run} to collect per-layer
performance data and generate deployment analysis materials. The
implementation of this reporting stage is still evolving and has not yet
been finalized, but it already provides a basis for subsequent
performance optimization and experience reuse.

\hypertarget{the-abstracting-role-of-agent-skills}{%
\subsection{The Abstracting Role of Agent
Skills}\label{the-abstracting-role-of-agent-skills}}

One key design of AIPC is to externalize deployment knowledge as
\textbf{Agent Skills}. These Skills do not replace the underlying SDK.
Instead, they provide recommended steps and execution order for the
agent, explicitly constrain which repair level should be addressed
first, and supply common failure patterns and suggested repair paths,
thereby reducing strategic drift in long deployment chains.

Therefore, AIPC's automation capability comes largely from the
combination of ``structured knowledge + tool capability + validation
loop,'' rather than from relying only on a stronger base model.

\hypertarget{a-generalized-preprocessing-view-across-multiple-backends}{%
\subsection{A Generalized Preprocessing View Across Multiple
Backends}\label{a-generalized-preprocessing-view-across-multiple-backends}}

Dynamic-shape fixing and operator replacement do not serve QAIRT alone;
they can also be viewed as model preprocessing steps before entering
other hardware-specific inference frameworks. Therefore, the
representation-layer parts of AIPC have some potential for cross-backend
migration. Although this paper remains implementation-focused on QAIRT,
its front-end workflow and methodological abstraction can in principle
be extended to other deployment targets such as TVM.

\hypertarget{experimental-design}{%
\section{Experimental Design}\label{experimental-design}}

\hypertarget{evaluation-goals}{%
\subsection{Evaluation Goals}\label{evaluation-goals}}

The experiments in this paper do not aim to build a large standardized
benchmark. Instead, they evaluate the following questions through case
studies:

\begin{enumerate}
\def\labelenumi{\arabic{enumi}.}
\tightlist
\item
  Can AIPC complete end-to-end deployment on representative models?
\item
  What are the primary automation barriers for models of different
  complexity?
\item
  How do different AI agents behave differently inside the same
  workflow?
\item
  Which deployment steps benefit most from automation, and which still
  require human leadership?
\end{enumerate}

\hypertarget{test-models}{%
\subsection{Test Models}\label{test-models}}

The following models are analyzed in this paper: - \textbf{ESRGAN}
{[}12{]} (image super-resolution): mainly convolutional, with a
relatively regular graph structure and few deployment obstacles, making
it suitable as a baseline case for automated workflow validation. -
\textbf{YOLOv8} {[}11{]} (object detection): mature structure and strong
ecosystem support. Its main deployment challenge lies in decoupling
postprocessing logic such as NMS and coordinate scaling. -
\textbf{LPRNet} {[}15{]} (license plate recognition): a lightweight
convolutional network whose main challenge is export-stage compatibility
around the \texttt{MaxPool3d} operator, requiring structural replacement
before successful conversion. - \textbf{YOLO-World} {[}16{]}
(open-vocabulary object detection): combines visual detection with text
features. Deployment challenges focus on replacing the \texttt{Einsum}
operator, handling multimodal interfaces, and aligning postprocessing
logic. - \textbf{YOLO26} {[}21{]} (next-generation object detection):
uses a newer network architecture whose toolchain compatibility is still
evolving. The main challenge lies in compatibility adaptation between
the exported graph and the QAIRT toolchain. - \textbf{Whisper} {[}13{]}
(speech recognition): uses a Transformer encoder-decoder structure. Its
main deployment difficulty lies in the inherent conflict between dynamic
sequence lengths, autoregressive token-by-token decoding, and the NPU
preference for fixed-shape batch computation. - \textbf{DeepSeek-R1}
{[}14{]} (extended observation): a large language model that raises
stronger demands in model size, dynamic decoding, operator support, and
resource constraints. It is not included in the full quantitative
comparison in this paper, but is discussed as an extended future
direction.

In the latest toolkit version, operator patching is supported by an
operator-patching database, making many cases easier to resolve. To
reproduce the original issue, the operator-patching database should be
removed.

\hypertarget{evaluated-agents}{%
\subsection{Evaluated Agents}\label{evaluated-agents}}

This paper observes several mainstream AI agent/model combinations.
Among them, the following three agent categories completed the full
evaluation workflow including inference validation, and their results
are included in the quantitative analysis in Section 6. Due to
test-environment resource constraints, this phase prioritized freely
available services. \textbf{All model information reflects the status
checked on March 30, 2026.}

\begin{longtable}[]{@{}lll@{}}
\toprule\noalign{}
Model Name & Context Window & Evaluation Platform \\
\midrule\noalign{}
\endhead
\bottomrule\noalign{}
\endlastfoot
Qwen 3.5 Coder & 32K tokens & Qwen CLI \\
MiniMax 2.5 {[}17{]} & 245K tokens & OpenCode \\
MiMo-V2-Pro & 1M tokens & OpenCode \\
\end{longtable}

The following agents participated in selected stages of the deployment
workflow. Due to test-environment constraints, full end-to-end
evaluation was not completed for this group; only partial results are
reported.

\begin{longtable}[]{@{}lll@{}}
\toprule\noalign{}
Model Name & Context Window & Evaluation Platform \\
\midrule\noalign{}
\endhead
\bottomrule\noalign{}
\endlastfoot
GPT-5.3 Codex & 128K tokens & Codex CLI \\
Gemini 3.1 Pro Preview & 1M tokens & Cline (VS Code) \\
Gemini 3.1 Flash Preview & 1M tokens & Cline (VS Code) \\
Claude Sonnet 4.6 & 200K tokens & Cline (VS Code) \\
\end{longtable}

All agents in this group were run in auto-approve (continuous execution)
mode, allowing uninterrupted workflow progression without per-step
confirmation. For agents executed via Cline (VS Code), API cost
estimates were read directly from Cline's built-in cost display in the
GUI.

It should be emphasized that the goal of this paper is not to produce a
strict ranking of general LLM agent capability, but to observe how
different agents differ in Skill adherence, recovery ability, and
task-completion stability within the same deployment workflow.

\hypertarget{evaluation-metrics}{%
\subsection{Evaluation Metrics}\label{evaluation-metrics}}

Since this paper is positioned as an engineering technical report, we
adopt metrics closer to deployment practice:

\begin{itemize}
\tightlist
\item
  \textbf{Number of manual interventions}: the number of points where
  human correction is required.
\item
  \textbf{End-to-end time}: approximate time from model preparation to
  successful validation of runnable output.
\item
  \textbf{Failure-mode types}: such as unsupported operators, path
  errors, environment misjudgment, workflow drift, and abnormal
  inference results.
\end{itemize}

\textbf{Notes on the experimental data}: Each row in the tables
corresponds to one independent run from scratch and records the actual
result of that run, rather than a statistical average over repeated
trials. The manual-intervention count records the number of times the
agent explicitly reported completion or failure and requested human
confirmation, or proactively requested human decision-making at key
points. End-to-end time is measured as the actual wall-clock time of the
full run, excluding model-weight download time but including agent
thinking, code generation, and command-execution wait time. Because this
stage relied mainly on free services, API-call cost is reported as an
empirical token-based estimate for reference only; actual costs may vary
with model version updates and error conditions.

\hypertarget{experimental-environment}{%
\subsection{Experimental Environment}\label{experimental-environment}}

\textbf{Hardware platform} - Development workstation: Qualcomm
Snapdragon X Elite2 (Windows 11) - Target device: Qualcomm Snapdragon X
Elite2 (Windows 11)

In the primary experiments in this paper, the development workstation
and target inference device are the same machine, both running on the
Snapdragon X Elite platform. This setup means that model conversion and
inference validation are completed in the same hardware environment,
eliminating the remote deployment step found in real cross-compilation
scenarios, but also making the performance data closer to ``local
inference'' rather than ``remote target-device inference.''

A second environment was also used for a limited set of tests,
representing a more realistic cross-compilation scenario:

\begin{itemize}
\tightlist
\item
  Development workstation: x86-64 Linux (Ubuntu 24.04)
\item
  Target device: Qualcomm RB8 development board (Ubuntu 24.04, ARM)
\end{itemize}

In this configuration, model conversion and ONNX-to-QAIRT build steps
are performed on the x86-64 host, while inference execution and
validation are carried out on the ARM target device. The agent is
responsible for the full workflow including artifact transfer, remote
environment setup, and on-device inference via SSH. This setup more
closely reflects real-world edge deployment practice. Due to environment
availability constraints, only a subset of models and agents were
evaluated in this configuration; results are reported separately where
applicable. Context Binary generation is optional and may be skipped in
this configuration.

\textbf{Software stack} - QAIRT SDK: 2.42 - QAI AppBuilder: 2.0

No additional specialized plugins were installed for the agents during
evaluation, in order to keep the environment as consistent as possible.

\hypertarget{execution-procedure}{%
\subsection{Execution Procedure}\label{execution-procedure}}

This paper uses a natural-language-prompt-driven, agent-executed,
human-observed evaluation method. A typical procedure includes the
following five phases:

\begin{itemize}
\tightlist
\item
  \textbf{Environment setup and baseline establishment}: Use the prompt
  \texttt{create\ example\ of\ {[}model{]}} to guide the agent to
  generate a runnable model example, and have a human confirm that
  baseline PyTorch inference works correctly.
\item
  \textbf{Create AIPC project}: Run \texttt{create\ aipc\ project}
  inside the model directory, letting the agent automatically create the
  project structure and generate the default configuration template.
\item
  \textbf{Configuration confirmation}: Use
  \texttt{show\ project\ configs} to inspect and confirm the current
  project configuration.
\item
  \textbf{Automated execution and manual intervention}: Use
  \texttt{do\ all\ work\ for\ the\ plan} to start the full deployment
  workflow; the developer observes execution and records intervention
  points if the workflow is interrupted or abnormal.
\item
  \textbf{Deployment report generation}: After completion, record the
  deployment analysis report generated by the agent, including
  conversion results, accuracy alignment, and key steps.
\end{itemize}

The experiments in this paper mainly use QNN FP16 mode. The quantization
stage requires appropriate calibration data and is not yet included in
the complete evaluation range.

\hypertarget{results-and-case-analysis}{%
\section{Results and Case Analysis}\label{results-and-case-analysis}}

This section analyzes the experimental results mainly through case-based
comparative observation. First, deployment behavior is examined model by
model from lower to higher complexity. Second, the AI agents involved in
the evaluation are compared laterally, with emphasis on workflow
adherence, failure recovery, and execution stability within the AIPC
workflow rather than on ranking model capability. The tables in this
section are intended to present representative case records and
failure-mode observations, and should not be interpreted as agent
rankings under strictly controlled conditions.

\hypertarget{analysis-by-deployment-model}{%
\subsection{Analysis by Deployment
Model}\label{analysis-by-deployment-model}}

This subsection presents representative case analysis in order of
increasing model complexity. Because the number of test runs differs
across models and agents, the focus is on representative behavior
patterns rather than direct horizontal comparison of single numbers.
Across the evaluated models, the main automation bottlenecks can be
broadly grouped into three categories: application-level pipeline
decoupling issues in otherwise regular vision models,
operator-compatibility and graph-rewrite issues in newer or multimodal
architectures, and structural mismatch between autoregressive decoding
patterns and NPU-oriented fixed-shape deployment assumptions.

\hypertarget{esrgan-high-automation-feasibility-on-a-standard-convolutional-model}{%
\subsubsection{ESRGAN: High Automation Feasibility on a Standard
Convolutional
Model}\label{esrgan-high-automation-feasibility-on-a-standard-convolutional-model}}

ESRGAN is one of the most stable models for automation in this study.
Its network structure is regular, its operator ecosystem is mature, and
its postprocessing is relatively simple, so AIPC can usually complete
the full ONNX-to-QNN workflow smoothly across most agents. The main
caveat is fixed input-shape configuration: if export parameters do not
match the test image size, inference may still run, but the results
become obviously incorrect. Once this constraint is explicitly stated,
the agent can usually correct it on its own.

\begin{longtable}[]{@{}
  >{\raggedright\arraybackslash}p{(\columnwidth - 8\tabcolsep) * \real{0.1667}}
  >{\raggedright\arraybackslash}p{(\columnwidth - 8\tabcolsep) * \real{0.1667}}
  >{\centering\arraybackslash}p{(\columnwidth - 8\tabcolsep) * \real{0.1667}}
  >{\centering\arraybackslash}p{(\columnwidth - 8\tabcolsep) * \real{0.3333}}
  >{\raggedright\arraybackslash}p{(\columnwidth - 8\tabcolsep) * \real{0.1667}}@{}}
\toprule\noalign{}
\begin{minipage}[b]{\linewidth}\raggedright
Agent
\end{minipage} & \begin{minipage}[b]{\linewidth}\raggedright
Backend
\end{minipage} & \begin{minipage}[b]{\linewidth}\centering
Time
\end{minipage} & \begin{minipage}[b]{\linewidth}\centering
Manual Interventions
\end{minipage} & \begin{minipage}[b]{\linewidth}\raggedright
Notes
\end{minipage} \\
\midrule\noalign{}
\endhead
\bottomrule\noalign{}
\endlastfoot
MiMo-V2-Pro & QNN & 9 min & 0 & \\
\multicolumn{5}{@{}p{\dimexpr\linewidth-2\tabcolsep\relax}@{}}{\textit{Note: Full workflow completed smoothly.}} \\
Qwen 3.5 Coder & QNN & 14 min & 0 & \\
\multicolumn{5}{@{}p{\dimexpr\linewidth-2\tabcolsep\relax}@{}}{\textit{Note: Full workflow completed smoothly.}} \\
MiniMax 2.5 & QNN & 7 min & 1 & \\
\multicolumn{5}{@{}p{\dimexpr\linewidth-2\tabcolsep\relax}@{}}{\textit{Note: The report was not generated; human retriggering was required.}} \\
\end{longtable}

\textbf{Remote inference on Linux (x86-64 host → ARM target, via SSH)}

\begin{longtable}[]{@{}
  >{\raggedright\arraybackslash}p{(\columnwidth - 8\tabcolsep) * \real{0.1667}}
  >{\raggedright\arraybackslash}p{(\columnwidth - 8\tabcolsep) * \real{0.1667}}
  >{\centering\arraybackslash}p{(\columnwidth - 8\tabcolsep) * \real{0.1667}}
  >{\centering\arraybackslash}p{(\columnwidth - 8\tabcolsep) * \real{0.3333}}
  >{\raggedright\arraybackslash}p{(\columnwidth - 8\tabcolsep) * \real{0.1667}}@{}}
\toprule\noalign{}
\begin{minipage}[b]{\linewidth}\raggedright
Agent
\end{minipage} & \begin{minipage}[b]{\linewidth}\raggedright
Backend
\end{minipage} & \begin{minipage}[b]{\linewidth}\centering
Time
\end{minipage} & \begin{minipage}[b]{\linewidth}\centering
Manual Interventions
\end{minipage} & \begin{minipage}[b]{\linewidth}\raggedright
Notes
\end{minipage} \\
\midrule\noalign{}
\endhead
\bottomrule\noalign{}
\endlastfoot
GPT-5.3 Codex & QNN & 13 min & 0 & \\
\multicolumn{5}{@{}p{\dimexpr\linewidth-2\tabcolsep\relax}@{}}{\textit{Note: Completed successfully.}} \\
Claude Sonnet 4.6 & QNN & 14 min & 0 & \\
\multicolumn{5}{@{}p{\dimexpr\linewidth-2\tabcolsep\relax}@{}}{\textit{Note: Estimated API cost: USD 2.2.}} \\
Gemini 3.1 Flash Preview & QNN & 20 min & 3 & \\
\multicolumn{5}{@{}p{\dimexpr\linewidth-2\tabcolsep\relax}@{}}{\textit{Note: Failed to switch to remote execution; additional errors occurred. Estimated API cost: USD 5.}} \\
\end{longtable}

This result suggests that, for models with regular structure and clear
interfaces, AIPC provides practical value within the current test range
and can handle much of the repetitive deployment work. One case also
illustrates a cost-amplification risk: when an initial error is not
properly recovered, subsequent cascading failures can significantly
inflate API costs --- in the observed instance, a single unresolved
deviation led to a chain of downstream errors that drove the total cost
to approximately USD 5, roughly 2--7× higher than successful runs on the
same model.

\hypertarget{yolov8-postprocessing-decoupling-is-critical-to-automation-success}{%
\subsubsection{YOLOv8: Postprocessing Decoupling Is Critical to
Automation
Success}\label{yolov8-postprocessing-decoupling-is-critical-to-automation-success}}

The main difficulty of YOLOv8 is not the backbone network itself, but
the fact that the official inference pipeline mixes in substantial
postprocessing logic, especially NMS and coordinate mapping. If the
agent treats the original flow as one monolithic unit, exported
inference often produces shifted bounding boxes or incorrect ROIs.
AIPC's modular design allows the agent to focus on repairing the
postprocessing part while preserving the stability of the main model.

Two additional evaluation settings were observed for YOLOv8 in order to
understand workflow robustness across environments.

\textbf{Early toolkit version without the operator-patching database
(Windows on ARM)}

\begin{longtable}[]{@{}
  >{\raggedright\arraybackslash}p{(\columnwidth - 8\tabcolsep) * \real{0.1667}}
  >{\raggedright\arraybackslash}p{(\columnwidth - 8\tabcolsep) * \real{0.1667}}
  >{\centering\arraybackslash}p{(\columnwidth - 8\tabcolsep) * \real{0.1667}}
  >{\centering\arraybackslash}p{(\columnwidth - 8\tabcolsep) * \real{0.3333}}
  >{\raggedright\arraybackslash}p{(\columnwidth - 8\tabcolsep) * \real{0.1667}}@{}}
\toprule\noalign{}
\begin{minipage}[b]{\linewidth}\raggedright
Agent
\end{minipage} & \begin{minipage}[b]{\linewidth}\raggedright
Backend
\end{minipage} & \begin{minipage}[b]{\linewidth}\centering
Time
\end{minipage} & \begin{minipage}[b]{\linewidth}\centering
Manual Interventions
\end{minipage} & \begin{minipage}[b]{\linewidth}\raggedright
Notes
\end{minipage} \\
\midrule\noalign{}
\endhead
\bottomrule\noalign{}
\endlastfoot
Qwen 3.5 Coder & QNN & 15 min & 0 & \\
\multicolumn{5}{@{}p{\dimexpr\linewidth-2\tabcolsep\relax}@{}}{\textit{Note: Completed successfully.}} \\
Qwen 3.5 Coder & SNPE & 6 min & 0 & \\
\multicolumn{5}{@{}p{\dimexpr\linewidth-2\tabcolsep\relax}@{}}{\textit{Note: Completed successfully.}} \\
MiMo-V2-Pro & QNN & 16 min & 0 & \\
\multicolumn{5}{@{}p{\dimexpr\linewidth-2\tabcolsep\relax}@{}}{\textit{Note: Completed successfully.}} \\
MiniMax 2.5 & QNN & 18 min & 2 & \\
\multicolumn{5}{@{}p{\dimexpr\linewidth-2\tabcolsep\relax}@{}}{\textit{Note: Misjudged hardware warnings and referenced the wrong SDK version.}} \\
MiniMax 2.5 & QNN & 12 min & 0 & \\
\multicolumn{5}{@{}p{\dimexpr\linewidth-2\tabcolsep\relax}@{}}{\textit{Note: Completed successfully after rerun.}} \\
\end{longtable}

\textbf{Remote inference on Linux (x86-64 host → ARM target, via SSH)}

\begin{longtable}[]{@{}
  >{\raggedright\arraybackslash}p{(\columnwidth - 8\tabcolsep) * \real{0.1667}}
  >{\raggedright\arraybackslash}p{(\columnwidth - 8\tabcolsep) * \real{0.1667}}
  >{\centering\arraybackslash}p{(\columnwidth - 8\tabcolsep) * \real{0.1667}}
  >{\centering\arraybackslash}p{(\columnwidth - 8\tabcolsep) * \real{0.3333}}
  >{\raggedright\arraybackslash}p{(\columnwidth - 8\tabcolsep) * \real{0.1667}}@{}}
\toprule\noalign{}
\begin{minipage}[b]{\linewidth}\raggedright
Agent
\end{minipage} & \begin{minipage}[b]{\linewidth}\raggedright
Backend
\end{minipage} & \begin{minipage}[b]{\linewidth}\centering
Time
\end{minipage} & \begin{minipage}[b]{\linewidth}\centering
Manual Interventions
\end{minipage} & \begin{minipage}[b]{\linewidth}\raggedright
Notes
\end{minipage} \\
\midrule\noalign{}
\endhead
\bottomrule\noalign{}
\endlastfoot
GPT-5.3 Codex & QNN & 8 min & 0 & \\
\multicolumn{5}{@{}p{\dimexpr\linewidth-2\tabcolsep\relax}@{}}{\textit{Note: Completed successfully.}} \\
Gemini 3.1 Flash Preview & QNN & 10 min & 0 & \\
\multicolumn{5}{@{}p{\dimexpr\linewidth-2\tabcolsep\relax}@{}}{\textit{Note: Estimated API cost: USD 0.46.}} \\
Gemini 3.1 Pro Preview & QNN & 20 min & 1 & \\
\multicolumn{5}{@{}p{\dimexpr\linewidth-2\tabcolsep\relax}@{}}{\textit{Note: One additional intervention was required to verify ROI detection results. Estimated API cost: USD 5.5.}} \\
Claude Sonnet 4.6 & QNN & 13 min & 0 & \\
\multicolumn{5}{@{}p{\dimexpr\linewidth-2\tabcolsep\relax}@{}}{\textit{Note: Estimated API cost: USD 2.53.}} \\
\end{longtable}

This case indicates that the key to deployment automation is not just
improving export success rate, but ensuring that application-layer logic
in the inference pipeline can be explicitly decomposed and validated
step by step.

\hypertarget{lprnet-the-value-of-operator-replacement-templates}{%
\subsubsection{LPRNet: The Value of Operator-Replacement
Templates}\label{lprnet-the-value-of-operator-replacement-templates}}

The typical obstacle in LPRNet is failure to export \texttt{MaxPool3d}.

\begin{longtable}[]{@{}
  >{\raggedright\arraybackslash}p{(\columnwidth - 8\tabcolsep) * \real{0.1667}}
  >{\raggedright\arraybackslash}p{(\columnwidth - 8\tabcolsep) * \real{0.1667}}
  >{\centering\arraybackslash}p{(\columnwidth - 8\tabcolsep) * \real{0.1667}}
  >{\centering\arraybackslash}p{(\columnwidth - 8\tabcolsep) * \real{0.3333}}
  >{\raggedright\arraybackslash}p{(\columnwidth - 8\tabcolsep) * \real{0.1667}}@{}}
\toprule\noalign{}
\begin{minipage}[b]{\linewidth}\raggedright
Agent
\end{minipage} & \begin{minipage}[b]{\linewidth}\raggedright
Backend
\end{minipage} & \begin{minipage}[b]{\linewidth}\centering
Time
\end{minipage} & \begin{minipage}[b]{\linewidth}\centering
Manual Interventions
\end{minipage} & \begin{minipage}[b]{\linewidth}\raggedright
Notes
\end{minipage} \\
\midrule\noalign{}
\endhead
\bottomrule\noalign{}
\endlastfoot
Qwen 3.5 Coder & QNN & 9 min & 0 & \\
\multicolumn{5}{@{}p{\dimexpr\linewidth-2\tabcolsep\relax}@{}}{\textit{Note: Completed successfully.}} \\
MiniMax 2.5 & QNN & 18 min & - & \\
\multicolumn{5}{@{}p{\dimexpr\linewidth-2\tabcolsep\relax}@{}}{\textit{Note: The first run did not complete because the workflow was interrupted.}} \\
MiniMax 2.5 & QNN & 8 min & 0 & \\
\multicolumn{5}{@{}p{\dimexpr\linewidth-2\tabcolsep\relax}@{}}{\textit{Note: The second run completed successfully after correction.}} \\
MiMo-V2-Pro & QNN & 20 min & 0 & \\
\multicolumn{5}{@{}p{\dimexpr\linewidth-2\tabcolsep\relax}@{}}{\textit{Note: Completed successfully.}} \\
\end{longtable}

This result indicates that summarizing common compatibility issues into
reusable repair patterns is a key way to improve stability in automated
deployment.

\hypertarget{yolo26-friction-between-new-architectures-and-evolving-toolchains}{%
\subsubsection{YOLO26: Friction Between New Architectures and Evolving
Toolchains}\label{yolo26-friction-between-new-architectures-and-evolving-toolchains}}

The YOLO26 evaluation focuses on AIPC's adaptability to newer model
structures. ONNX export of this model succeeds smoothly, and the QNN
conversion stage can also generate intermediate artifacts, but Context
Binary compilation may still encounter unsupported operators, requiring
operator patching or fallback to the SNPE DLC path to complete
deployment.

In this case, the typical problematic operator is \texttt{Mod}.
\texttt{Mod(a,\ b)} is equivalent to \texttt{a\ \%\ b}, and in theory
can be expanded into \texttt{a\ -\ b\ *\ floor(a\ /\ b)}. However,
\texttt{floor} is also unsupported during QNN conversion, so the entire
expression must be further rewritten as a pure basic-operator
composition of \texttt{Div\ +\ Mul\ +\ Sub} before conversion succeeds.
This two-level replacement chain is the stage where differences in agent
behavior are most obvious.

\begin{longtable}[]{@{}
  >{\raggedright\arraybackslash}p{(\columnwidth - 8\tabcolsep) * \real{0.1667}}
  >{\raggedright\arraybackslash}p{(\columnwidth - 8\tabcolsep) * \real{0.1667}}
  >{\centering\arraybackslash}p{(\columnwidth - 8\tabcolsep) * \real{0.1667}}
  >{\centering\arraybackslash}p{(\columnwidth - 8\tabcolsep) * \real{0.3333}}
  >{\raggedright\arraybackslash}p{(\columnwidth - 8\tabcolsep) * \real{0.1667}}@{}}
\toprule\noalign{}
\begin{minipage}[b]{\linewidth}\raggedright
Agent
\end{minipage} & \begin{minipage}[b]{\linewidth}\raggedright
Backend
\end{minipage} & \begin{minipage}[b]{\linewidth}\centering
Time
\end{minipage} & \begin{minipage}[b]{\linewidth}\centering
Manual Interventions
\end{minipage} & \begin{minipage}[b]{\linewidth}\raggedright
Notes
\end{minipage} \\
\midrule\noalign{}
\endhead
\bottomrule\noalign{}
\endlastfoot
MiMo-V2-Pro & QNN & 12 min & 0 & \\
\multicolumn{5}{@{}p{\dimexpr\linewidth-2\tabcolsep\relax}@{}}{\textit{Note: Completed successfully.}} \\
Qwen 3.5 Coder & QNN & 120 min & 1 & \\
\multicolumn{5}{@{}p{\dimexpr\linewidth-2\tabcolsep\relax}@{}}{\textit{Note: The first run failed; completion required human guidance and repair.}} \\
Qwen 3.5 Coder & QNN & 25 min & 1 & \\
\multicolumn{5}{@{}p{\dimexpr\linewidth-2\tabcolsep\relax}@{}}{\textit{Note: It repeatedly switched to SNPE automatically and did not complete under QNN without human guidance.}} \\
Qwen 3.5 Coder & QNN & 53 min & 3 & \\
\multicolumn{5}{@{}p{\dimexpr\linewidth-2\tabcolsep\relax}@{}}{\textit{Note: It completed with SNPE first, and then completed under the intended flow after AIPC Skill use was enforced and MiniMax's operator-replacement strategy was referenced.}} \\
MiniMax 2.5 & QNN & 29 min & 1 & \\
\multicolumn{5}{@{}p{\dimexpr\linewidth-2\tabcolsep\relax}@{}}{\textit{Note: It misjudged the ARM64/x86\_64 emulation relationship.}} \\
\end{longtable}

The YOLO26 results suggest that human intervention is still needed in
some cases to match expected deployment goals.

\hypertarget{yolo-world-structural-bottlenecks-in-complex-multimodal-models}{%
\subsubsection{YOLO-World: Structural Bottlenecks in Complex Multimodal
Models}\label{yolo-world-structural-bottlenecks-in-complex-multimodal-models}}

YOLO-World combines the postprocessing complexity of detection with
cross-modal operator issues in open-vocabulary models. \texttt{Einsum}
is the most typical incompatible operator in this model and must be
replaced before export. In addition, path management, text-embedding
preprocessing, and result interpretation further increase the chance of
agent error.

\begin{longtable}[]{@{}
  >{\raggedright\arraybackslash}p{(\columnwidth - 8\tabcolsep) * \real{0.1667}}
  >{\raggedright\arraybackslash}p{(\columnwidth - 8\tabcolsep) * \real{0.1667}}
  >{\centering\arraybackslash}p{(\columnwidth - 8\tabcolsep) * \real{0.1667}}
  >{\centering\arraybackslash}p{(\columnwidth - 8\tabcolsep) * \real{0.3333}}
  >{\raggedright\arraybackslash}p{(\columnwidth - 8\tabcolsep) * \real{0.1667}}@{}}
\toprule\noalign{}
\begin{minipage}[b]{\linewidth}\raggedright
Agent
\end{minipage} & \begin{minipage}[b]{\linewidth}\raggedright
Backend
\end{minipage} & \begin{minipage}[b]{\linewidth}\centering
Time
\end{minipage} & \begin{minipage}[b]{\linewidth}\centering
Manual Interventions
\end{minipage} & \begin{minipage}[b]{\linewidth}\raggedright
Notes
\end{minipage} \\
\midrule\noalign{}
\endhead
\bottomrule\noalign{}
\endlastfoot
MiMo-V2-Pro & QNN & 25 min & 0 & \\
\multicolumn{5}{@{}p{\dimexpr\linewidth-2\tabcolsep\relax}@{}}{\textit{Note: Completed successfully.}} \\
Qwen 3.5 Coder & QNN & 32 min & 1 & \\
\multicolumn{5}{@{}p{\dimexpr\linewidth-2\tabcolsep\relax}@{}}{\textit{Note: A path error caused temporary drift toward an incorrect alternative approach.}} \\
MiniMax 2.5 & QNN & 113 min & 0 & \\
\multicolumn{5}{@{}p{\dimexpr\linewidth-2\tabcolsep\relax}@{}}{\textit{Note: The long runtime was likely affected by workflow drift.}} \\
\end{longtable}

This case shows that once model structure goes beyond a single-modal
convolutional network, the main pressure on automation is no longer just
operator support, but maintaining strategy stability when multiple
classes of problems appear at once.

\hypertarget{whisper-deep-conflict-between-model-structure-and-hardware-execution-mode}{%
\subsubsection{Whisper: Deep Conflict Between Model Structure and
Hardware Execution
Mode}\label{whisper-deep-conflict-between-model-structure-and-hardware-execution-mode}}

Whisper's encoder-decoder structure is far more complex than that of
convolutional models. The natural conflict between autoregressive
token-by-token decoding with dynamic sequence length and the NPU
preference for fixed-shape batch computation means that Whisper
deployment is not merely a question of operator support from the
beginning.

In current tests, the encoder can be migrated to NPU execution, but the
decoder often falls back to CPU execution because of dynamic shapes and
autoregressive dependencies. This is not a satisfactory outcome, but it
reveals the core issue: there is a structural mismatch between the
decoder's execution style and the NPU's compilation assumptions.

Preliminary exploration suggests that if the decoder is manually
statically reshaped at the PyTorch level and the KV Cache interface is
exposed explicitly, the decoder may also be brought onto the NPU
conversion path, enabling more complete NPU execution coverage. This
direction is still in an early exploratory stage, and the experimental
results have not yet been systematically organized, but tests with
MiMo-V2-Pro have already shown some feasibility.

This case also points to a future expansion boundary for AIPC. At
present, AIPC's model surgery mainly focuses on operator-compatibility
repair. For models such as Whisper, however, the real bottleneck lies in
\textbf{model-structure transformation for NPU execution patterns},
including static reshaping, explicit KV Cache exposure, and
decoding-loop refactoring. These changes go beyond simple operator
replacement and require the agent to have a deeper understanding of
model execution semantics and proactive restructuring for target
hardware. Systematically incorporating this kind of ``NPU-oriented model
redesign'' into the AIPC workflow is an important future research
direction.

\hypertarget{deepseek-r1-and-larger-models-extended-directions}{%
\subsubsection{DeepSeek-R1 and Larger Models: Extended
Directions}\label{deepseek-r1-and-larger-models-extended-directions}}

Large language models such as DeepSeek-R1 impose stronger requirements
in parameter scale, dynamic decoding, KV cache handling, and operator
coverage. At this stage, AIPC can provide workflow structure and limited
local automation for such models, but cannot yet complete deployment
reliably without significant human participation. Notably, compared with
the accuracy-alignment challenges seen in Whisper, large language models
may be relatively easier in precision handling. This direction will be
explored further in future work.

\hypertarget{ai-agent-behavior-analysis}{%
\subsection{AI Agent Behavior
Analysis}\label{ai-agent-behavior-analysis}}

Based on multiple rounds of case observation, the behavior of the
evaluated agents within the AIPC workflow can be qualitatively
summarized as follows, with emphasis on workflow adherence, failure
recovery, and execution stability:

\begin{itemize}
\tightlist
\item
  \textbf{Qwen 3.5 Coder}: Overall stable in standard workflows and
  usually able to follow Skill constraints step by step.
\item
  \textbf{MiniMax 2.5}: Improved relative to earlier versions, but in
  some cases still more prone to workflow drift, misinterpreting
  hardware warnings, or bypassing Skills to directly invoke low-level
  commands, increasing runtime and result variance.
\item
  \textbf{MiMo-V2-Pro}: Showed smooth execution in multiple cases,
  relatively high adherence to Skill constraints, and
  zero-manual-intervention completion on several structurally regular
  models.
\end{itemize}

From these case observations, differences in \textbf{adherence to Skill
constraints} appear to matter more for deployment-task quality than
differences in generic coding ability. This supports an important
conclusion: in deployment tasks, an agent's ability to follow the
workflow is as important as its platform knowledge. A medium-capability
agent that follows process strictly is often more suitable for
engineering automation than a stronger but frequently drifting one.

In addition, \textbf{GPT-5.3 (Codex CLI)}, \textbf{Claude Sonnet 4.6},
and \textbf{Gemini 3.1 Flash / Pro} have so far only been observed
through the cross-compilation stage because of environment limitations.
Their overall usage flow was smooth and usable, and complete end-to-end
inference validation will be added when conditions permit.

This paper also explored using lighter locally deployable models such as
\textbf{gpt-oss-120b}, \textbf{Qwen-Next-80b}, and \textbf{Gemma 4} as
the agent-driving layer. However, under the current environment, none of
these models could stably complete functional verification within the
AIPC workflow, and they often experienced workflow interruptions. We
believe local small models as agents have potential value in privacy
protection and lower call cost, but their instruction-following ability
and tool-use stability in complex deployment tasks still require further
study and optimization.

\hypertarget{typical-failure-modes}{%
\subsubsection{Typical Failure Modes}\label{typical-failure-modes}}

Based on repeated case observations, this paper identifies several
recurring failure modes in agent execution. Importantly, these failures
are not caused by agents lacking the ability to solve
operator-compatibility problems. In fact, with enough attempts, most
agents can eventually find valid replacement or repair strategies. The
real issue is that missing workflow constraints, execution-path
deviation, or environment misjudgment prevent that capability from being
applied in the right direction.

\begin{itemize}
\item
  \textbf{Ignoring Skill constraints and falling into repeated trial and
  error}: Some agents, when facing conversion failure, did not follow
  the repair path defined by the Skill and instead tried many
  alternatives on their own. Although the agent itself was capable of
  finding a correct solution, each attempt lacked a clear validation
  exit, errors accumulated, and the agent eventually stopped after
  exhausting retries instead of returning to the correct repair level
  and restarting. The essence of this mode is not ``cannot do it,'' but
  ``was not guided in the right direction.''
\item
  \textbf{Applying operator patches by writing ONNX files to disk and
  then giving up}: AIPC requires operator patches to be done through
  in-memory graph manipulation to preserve reproducibility and
  traceability. However, some agents chose to write modified ONNX files
  directly back to disk and use them as inputs to later steps, causing
  the patch logic to become decoupled from the original model and making
  intermediate states hard to reproduce. Worse, when this approach
  encountered later conversion failures, the agent often could not
  localize the root cause and eventually gave up instead of continuing
  to repair. The root problem is noncompliance with the AIPC operation
  convention, not lack of technical ability for operator replacement.
\item
  \textbf{Misjudging runtime platform architecture}: Under Windows on
  ARM, some agents incorrectly judged the current environment as x86\_64
  and therefore invoked the wrong toolchain libraries or generated build
  artifacts mismatched to the target architecture. This is a form of
  ``silent failure'': the agent does not fail immediately, and the
  problem only appears later at inference or linking time, while the
  error messages often do not directly point back to the architecture
  misjudgment. The issue here is not lack of cross-architecture handling
  ability, but an incorrect premise formed during environment perception
  that poisons all subsequent reasoning.
\item
  \textbf{Premature abandonment caused by ambiguous prompts or code
  definitions}: When interface definitions are ambiguous, agents easily
  form incorrect assumptions. For example, the ONNX Wrapper tool
  initially implemented only the QNN backend and later added SNPE
  support, but the class and function names were not updated
  accordingly. When an agent invoked the tool along the SNPE path and
  encountered an execution failure, it often attributed the failure to
  ``SNPE not supporting the feature'' rather than tracing it back to
  ambiguity in the wrapper interface definition, and then chose to stop
  instead of continuing to investigate. This failure mode shows that
  interface naming and documentation clarity directly affect whether the
  agent can correctly identify root causes.
\end{itemize}

\hypertarget{engineering-lessons-and-limitations}{%
\section{Engineering Lessons and
Limitations}\label{engineering-lessons-and-limitations}}

\hypertarget{engineering-lessons}{%
\subsection{Engineering Lessons}\label{engineering-lessons}}

According to the practice in this paper, AIPC works mainly because of
the following factors:

\begin{itemize}
\tightlist
\item
  \textbf{Modularity comes before end-to-end monoliths}: Splitting
  preprocessing, inference, and postprocessing is the basis for
  improving automation success rate. Many errors do not come from the
  model itself, but from surrounding logic that was not migrated
  consistently.
\item
  \textbf{Skill packaging comes before free-form prompting}: For long
  deployment chains, natural-language prompting alone is not enough to
  constrain the agent. Organizing knowledge into Skills, script
  templates, and error references significantly reduces strategic drift.
\item
  \textbf{Every step needs a verifiable exit}: If the agent works too
  long in an unverifiable state, it will easily accumulate errors and
  drift away from the task goal. Golden outputs, shape checks, path
  checks, and numerical comparisons should therefore run through the
  entire workflow.
\item
  \textbf{Human intervention should be treated as data, not failure}:
  Automated deployment does not mean completely eliminating human
  participation. A more realistic approach is to review intervention
  points after each task, ask the AI for analysis, and let engineers
  decide whether to incorporate the corresponding repair path into the
  Skill or workflow template, selectively updating the workflow and
  continuously improving future automation.
\end{itemize}

\hypertarget{current-limitations}{%
\subsection{Current Limitations}\label{current-limitations}}

At present, AIPC still has several clear limitations:

\begin{itemize}
\tightlist
\item
  \textbf{Less-supported operators still depend on expert knowledge}:
  When a model contains new operators for which no mature equivalent
  replacement exists, the agent is unlikely to invent the correct
  implementation on its own, and expert guidance is still needed.
\item
  \textbf{Agent instruction following remains unstable}: Even agents
  with strong coding ability may ignore Skill constraints and jump into
  low-level command trial and error, increasing execution time and
  making outcomes unpredictable.
\item
  \textbf{Environment complexity exceeds pure text reasoning}: For
  example, the dual-emulation environment of Windows on ARM is complex
  even for human engineers, and it is even more likely to cause silent
  errors for agents. These problems require stronger environment
  detection and better tool feedback mechanisms.
\item
  \textbf{Current results are still mainly case observations}: The
  results in this paper are based primarily on case analysis from real
  engineering tasks rather than large-scale benchmarks with strictly
  controlled variables. Therefore, the conclusions are better understood
  as evidence of method feasibility and engineering experience than as
  absolute rankings of agent capability.
\end{itemize}

\hypertarget{future-work}{%
\section{Future Work}\label{future-work}}

\hypertarget{short-term-directions}{%
\subsection{Short-Term Directions}\label{short-term-directions}}

\begin{itemize}
\tightlist
\item
  \textbf{Context engineering and Skill-structure optimization}: The
  next stage will continue analyzing why agents deviate from the
  workflow, refining context organization, prompt hierarchy, and
  failure-handling rules, while also improving the execution harness,
  step constraints, and output conventions.
\item
  \textbf{Error knowledge base}: We plan to build a unified error
  knowledge base so that, upon failure, the agent can first query known
  error patterns and then execute the repair. After task completion, new
  errors and repair paths can be written back into the knowledge base,
  forming a continuously evolving mechanism of ``error occurs
  -\textgreater{} query knowledge base -\textgreater{} apply repair
  -\textgreater{} validate result -\textgreater{} write back new
  knowledge.''
\item
  \textbf{Support for more model families}: Future work will further
  explore edge-side automated deployment paths for Gemma/Qwen language
  models as well as models such as Stable Diffusion and PaddleOCR, in
  order to evaluate the cross-task generalization of AIPC.
\item
  \textbf{Optimization-chain extension}: Beyond quantization, AIPC will
  attempt to integrate pruning, mixed precision, and structured
  compression so that automation can move from ``deployable'' to
  ``better after deployment.''
\end{itemize}

\hypertarget{mid--and-long-term-directions}{%
\subsection{Mid- and Long-Term
Directions}\label{mid--and-long-term-directions}}

\begin{itemize}
\tightlist
\item
  \textbf{Cross-backend extension}: The front-end method of AIPC is not
  limited to QAIRT and can be extended to other target inference
  frameworks.
\item
  \textbf{Hierarchical agents and local-small-model collaboration}:
  Complex deployment currently still depends mainly on high-capability
  cloud models. In the future, one can explore letting lightweight local
  models handle standardized steps while falling back to cloud models
  for complex repairs, balancing cost, privacy, and success rate.
\item
  \textbf{Infrastructure for ``AI deploying AI''}: At a broader level,
  the workflow organized by AIPC is highly similar to a compiler pass
  pipeline. The deployment chain can be understood as an ordered
  sequence of staged transformations, from a PyTorch graph to ONNX, then
  to QAIRT intermediate representation, and finally to a Context Binary,
  with each step carrying specific transformation semantics and
  validation constraints. In the future, deployment infrastructure may
  evolve into an automated pass system partially driven by AI, gradually
  turning deployment from an experience-intensive manual activity into
  composable, verifiable, and continuously evolving software
  infrastructure.
\end{itemize}

\hypertarget{conclusion}{%
\section{Conclusion}\label{conclusion}}

This paper presents AIPC (AI Porting Conversion) as a technical report
on AI agent-driven automation for edge AI model deployment, with
Qualcomm AI Runtime (QAIRT) as the core implementation scenario.

AIPC is designed not as a fully autonomous deployment agent, but as a
constrained and recoverable workflow framework built around Agent
Skills, auxiliary scripts, and stage-wise validation. Within the cases
covered in this report, it can already automate a substantial portion of
deployment work for structurally regular vision models, significantly
shortening the engineering path to a runnable deployment result,
lowering the expertise barrier, and reducing repetitive manual effort.
For more complex models, AIPC does not yet eliminate the need for human
guidance, but it still provides practical support for execution, failure
localization, and bounded repair, thereby improving deployment
efficiency under realistic engineering constraints.

More broadly, this work suggests that the value of LLM agents in
deployment lies less in unrestricted autonomy than in reliable execution
within an explicitly engineered workflow. In this sense, AIPC provides a
practical starting point for ``AI deploying AI'': not as a claim of
complete autonomy today, but as an emerging form of engineering
infrastructure that can progressively reduce time, labor, and deployment
cost as error knowledge accumulates, Skill structures improve, and
backend coverage expands.

\hypertarget{acknowledgments}{%
\section{Acknowledgments}\label{acknowledgments}}

This work is an auxiliary study produced in the course of a
project-driven effort. We thank our colleagues for their support, which
made it possible to complete this work under tight schedules and limited
time resources. We also thank former colleagues Ioannis Nousias, Mark
Muir, and Adrian Giura, whose long-term collaboration style provided
important engineering inspiration for this work.

We also thank the relevant LLM tools and platforms for providing
evaluation resources that enabled the multi-agent case study. Part of
the practice in this paper was built on the open-source QAI AppBuilder
project, and the resulting work has also been contributed back to the
community in open-source form. We hope to continue improving the edge AI
deployment automation ecosystem together with more developers.

\hypertarget{references}{%
\section{References}\label{references}}

{[}1{]} NVIDIA Corporation. NVIDIA TensorRT: Programmable Inference
Accelerator. URL: https://developer.nvidia.com/tensorrt (accessed March
30, 2026).

{[}2{]} Intel Corporation. OpenVINO Toolkit: Open Visual Inference and
Neural Network Optimization. URL: https://docs.openvino.ai/ (accessed
March 30, 2026).

{[}3{]} Rockchip Electronics. RKNPU2: Rockchip Neural Processing Unit
SDK. URL: https://github.com/rockchip-linux/rknpu2 (accessed March 30,
2026).

{[}4{]} Anthropic. Claude Code: Agentic Coding in the Terminal. URL:
https://www.anthropic.com/claude-code (accessed March 30, 2026).

{[}5{]} GitHub. GitHub Copilot: AI Pair Programmer. URL:
https://github.com/features/copilot (accessed March 30, 2026).

{[}6{]} Anthropic. Claude Code Documentation. URL:
https://docs.anthropic.com/en/docs/claude-code (accessed March 30,
2026).

{[}7{]} Cursor. Cursor: The AI-first Code Editor. URL:
https://cursor.sh/ (accessed March 30, 2026).

{[}8{]} Qualcomm Technologies, Inc.~Qualcomm AI Runtime (QAIRT) SDK
Documentation. URL:
https://developer.qualcomm.com/software/qualcomm-ai-runtime (accessed
March 30, 2026).

{[}9{]} Linux Foundation. ONNX: Open Neural Network Exchange. URL:
https://onnx.ai/ (accessed March 30, 2026).

{[}10{]} Adam Paszke et al.~PyTorch: An imperative style,
high-performance deep learning library. In: Advances in Neural
Information Processing Systems 32 (2019), pp.~8026-8037.

{[}11{]} Ultralytics. Explore Ultralytics YOLOv8. URL:
https://docs.ultralytics.com/models/yolov8/ (accessed March 30, 2026).

{[}12{]} Xintao Wang et al.~ESRGAN: Enhanced super-resolution generative
adversarial networks. In: Proceedings of the European Conference on
Computer Vision Workshops (ECCVW). 2018.

{[}13{]} Alec Radford et al.~Robust speech recognition via large-scale
weak supervision. In: Proceedings of the 40th International Conference
on Machine Learning (ICML). 2023, pp.~28492-28518.

{[}14{]} DeepSeek-AI et al.~DeepSeek-R1: Incentivizing reasoning
capability in LLMs via reinforcement learning. 2025. arXiv: 2501.12948
{[}cs.CL{]}.

{[}15{]} Sergey Zherzdev and Alexey Gruzdev. LPRNet: License Plate
Recognition via Deep Neural Networks. 2018. arXiv: 1806.10447
{[}cs.CV{]}.

{[}16{]} Tao Cheng et al.~YOLO-World: Real-time open-vocabulary object
detection. In: Proceedings of the IEEE/CVF Conference on Computer Vision
and Pattern Recognition (CVPR). 2024.

{[}17{]} MiniMax. MiniMax: Large Language Model Platform. URL:
https://www.minimax.io/ (accessed March 30, 2026).

{[}18{]} SST. OpenCode: Open Source AI Coding Agent. URL:
https://github.com/sst/opencode (accessed March 30, 2026).

{[}19{]} Cline. Cline: Autonomous Coding Agent for VS Code. URL:
https://cline.bot/ (accessed March 30, 2026).

{[}20{]} OpenAI. Codex CLI: Lightweight Coding Agent in the Terminal.
URL: https://github.com/openai/codex (accessed March 30, 2026).

{[}21{]} Rijan Sapkota et al.~YOLO26: Key Architectural Enhancements and
Performance Benchmarking for Real-Time Object Detection. 2025. arXiv:
2509.25164 {[}cs.CV{]}.

{[}22{]} Tianqi Chen et al.~TVM: An automated end-to-end optimizing
compiler for deep learning. In: Proceedings of the 13th USENIX Symposium
on Operating Systems Design and Implementation (OSDI). 2018,
pp.~578-594.

{[}23{]} Chris Lattner et al.~MLIR: Scaling compiler infrastructure for
domain specific computation. In: Proceedings of the IEEE/ACM
International Symposium on Code Generation and Optimization (CGO). 2021,
pp.~2-14.

\end{document}